# Offcut-related step-flow and growth rate enhancement during (100) β-Ga$_2$O$_3$ homoepitaxy by metal-exchange catalyzed molecular beam epitaxy (MEXCAT-MBE)

P. Mazzolini[1(^),*], A. Falkenstein[2], Z. Galazka[3], M. Martin[2], O. Bierwagen[1*]

[1] *Paul-Drude-Institut für Festkörperelektronik Leibniz-Institut im Forschungsverbund, Hausvogteiplatz 5-7, 10117 Berlin, Germany*

[2] *Institute of Physical Chemistry, RWTH Aachen University, D-52056 Aachen, Germany*

[3] *Leibniz-Institut für Kristallzüchtung, Max-Born-Str. 2, 12489 Berlin, Germany*

(^) *Present address: Department of Mathematical, Physical and Computer Sciences, University of Parma, Viale delle Scienze 7/A, 43124 Parma, Italy*

*Email: piero.mazzolini@unipr.it – bierwagen@pdi-berlin.de*

In this work we investigate the growth of β-Ga$_2$O$_3$ homoepitaxial layers on top of (100) oriented substrates via indium-assisted metal exchange catalyzed molecular beam epitaxy (MEXCAT-MBE) which have exhibited prohibitively low growth rates by non-catalyzed MBE in the past. We demonstrate that the proper tuning of the MEXCAT growth parameters and the choice of a proper substrate offcut allow for the deposition of thin films with high structural quality via step-flow growth mechanism at relatively high growth rates for β-Ga$_2$O$_3$ homoepitaxy (*i.e.*, around 1.5 nm/min, ≈ 45% incorporation of the incoming Ga flux), making MBE growth on this orientation feasible. Moreover, through the employment of the investigated four different (100) substrate offcuts along the $[00\bar{1}]$ direction (*i.e.*, 0°, 2°, 4°, 6°) we give experimental evidence on the fundamental role of the $(\bar{2}01)$ step edges as nucleation sites for growth of (100)-oriented Ga$_2$O$_3$ films by MBE.

## THE MANUSCRIPT

β-Ga$_2$O$_3$ plays an important role for new generation power electronic devices.[1] A great advantage for this ultra-wide bandgap oxide is the possibility to produce bulk material from the melt,[2] enabling the homoepitaxial growth of high quality β-Ga$_2$O$_3$ thin films. Among the different growth techniques, molecular beam epitaxy (MBE)[3–5] and metalorganic chemical vapor deposition or metalorganic vapor phase epitaxy (MOCVD or MOVPE)[6–8] have so far emerged for providing the highest quality β-Ga$_2$O$_3$ homoepitaxial layers. Nonetheless, the chosen substrate orientation for homoepitaxial deposition of β-Ga$_2$O$_3$ can pose two major challenges: (1) the possible formation of structural defects and (2) the presence of different growth rates (GRs). The first point is related to the low symmetry of the monoclinic cell of β-Ga$_2$O$_3$ which, due to the possible double positioning of Ga atoms and island coalescence during the growth process, eventually results in the creation of twins in both the (100)[9,10] and $(\bar{2}01)$[4] homoepitaxial growth which can affect the electrical properties of the layers.[11] Nonetheless, for MOVPE (100) homoepitaxy it has been possible to overcome this problem with the choice of an appropriate substrate offcut so to allow the step flow growth of the layers in presence of ($\bar{2}$01) steps.[7,8] The second challenge in β-Ga$_2$O$_3$ homoepitaxy is the different GR recorded as a function of different substrate orientations. This is mostly affecting the MBE growth technique[3,12] and is mostly related to the peculiar growth kinetics of Ga$_2$O$_3$ during MBE growth which is involving the intermediate formation of the Ga$_2$O volatile suboxide;[13–15] the Ga$_2$O tendency to desorb from the sample surface before its further oxidation in Ga$_2$O$_3$ is dependent on the growth surface itself. This has so far limited the investigation of MBE grown β-Ga$_2$O$_3$ homoepitaxial layers to the (010) growth plane, *i.e.*, the one with the highest surface free energy;[7,16] for such orientation, GRs as high as ≈ 2.2 nm/min [3] and 3.2 nm/min [17] were reported for ozone and plasma-assisted MBE respectively. The other β-Ga$_2$O$_3$ surfaces and especially the most stable (100) cleavage plane[7,16] generally suffer from lower GRs in MBE (*e.g.*, ≈ 0.15 nm/min [3] and ≈ 0.3 nm/min [18] for ozone and plasma assisted MBE respectively). Differently, for MOVPE negligible substrate orientation dependencies have been reported with respect to the GR, probably because of a different Ga$_2$O$_3$ growth kinetics compared to the MBE process;[8] in particular for (100) MOVPE homoepitaxy high quality layers were deposited with GRs between 1.6 nm/min and 4.3 nm/min on substrates offcut towards [00$\bar{1}$].[8]

The (100) β-Ga$_2$O$_3$ growth surface is potentially very interesting because of the possibility to obtain step-flow growth resulting in smooth layers with high structural quality[7,8] [similar to what has been recently demonstrated in (010) homoepitaxy with proper substrate offcuts[19]]. Unfortunately, the low GRs of the MBE process so far have practically limited the homoepitaxial synthesis on the (100) orientation solely to the MOVPE technique. Nonetheless, it has been recently demonstrated that the addition of an In[20] or Sn[21] metal flux can allow to widen the growth window of Ga$_2$O$_3$ in MBE by the metal-exchange catalysis (MEXCAT) mechanism.[15] Indium-mediated MEXCAT has been recently successfully applied in the MBE homoepitaxial growth of β-Ga$_2$O$_3$ layers over different orientations,[4,5,19,22] as well as β-(Al,Ga)$_2$O$_3$ layers on β-Ga$_2$O$_3$(010) substrates.[23] In particular, with MEXCAT-MBE it has been possible to demonstrate that the use of (100) substrates with a 6° offcut results in homoepitaxial layers with a high structural quality comparable to the one obtained in MOVPE growths.[4,7] Nonetheless, the GR has been found to be still a function of the β-Ga$_2$O$_3$ growth surface,[4] proving that MEXCAT-MBE process can mitigate but not fully eliminate the partial loss of the incoming Ga-flux from highly stable surfaces like the (100) one. In particular, the GR obtained for high quality MEXCAT-MBE β-Ga$_2$O$_3$ homoepitaxial layers on 6°-off (100) substrates was around 0.27 nm/min (*i.e.*, correspondent to less than 10% incorporation of the incoming Ga-flux), the lowest one with respect to the other tested β-Ga$_2$O$_3$ orientations [*i.e.*, (010), (001), and ($\bar{2}$01)]. As a comparison, the same Ga-flux can be fully incorporated in (010)-oriented MEXCAT-MBE homoepitaxy (*i.e.*, 3.3 nm/min),[19] while Mauze *et al.*[5] with a similar experimental approach were able to obtain GRs as high as ≈ 5 nm/min on the same substrate orientation proving that the GR can be maximized by properly increasing the provided metal and oxygen fluxes.

In the present work, throughout the optimization of the MEXCAT-MBE deposition process and the understanding of the role of the offcut angle in (100) homoepitaxy, we demonstrate step-flow growth rates GRs of ≈ 1.5 nm/min (≈ 45% incorporation of Ga-flux) comparable to the ones obtained by MOVPE growth technique on offcut (100) substrates.

Mg-doped (100) β-$Ga_2O_3$ substrates with [00$\bar{1}$]-oriented offcuts of 0°, 2°, 4°, and 6° prepared from bulk crystals obtained by Czochralski method[24,25] were employed. The substrate preparation prior to the deposition is explained in our previous work.[4] The depositions were performed in an MBE chamber with an O-plasma source run at a power of 300 W for a fixed deposition time of 30 minutes. The beam equivalent pressure (BEP) of Ga was fixed to $BEP_{Ga}$ ≈ 1.2×10$^{-7}$ mbar (particle flux $\Phi_{Ga}$ = 2.2 nm$^{-2}$s$^{-1}$), corresponding to a GR ≈ 3.3 nm/min at full Ga-incorporation (thickness ≈ 100 nm). The In-flux for the MEXCAT process is set to be $\Phi_{In}$ = 1/3 $\Phi_{Ga}$ ($BEP_{In}$ ≈ 5.2×10$^{-8}$ mbar). Both the metal fluxes were maintained constant for all the growths, while the oxygen flow and the growth temperatures were varied among 0.75 standard cubic centimetres per minute (sccm) and 1 sccm, and $T_g$ = 700 – 800 °C. An $(Al_xGa_{1-x})_2O_3$ marker layer at the substrate-layer interface was deposited (deposition time = 80 s) in almost all the growth runs; the related details are reported in a previous work.[4] The (100) β-$Ga_2O_3$ substrates with different offcuts were co-loaded so that they are subjected to the same deposition conditions in a single growth run. The growth rate was calculated from the growth time and layer thickness determined using the In signal from time-of-flight secondary ion mass spectrometry (ToF-SIMS IV from IONTOF GmbH) depth profiling and/or by X-ray diffraction (XRD) fringes interspace in the vicinity of the β-$Ga_2O_3$ 400 reflex (out-of-plane 2θ-ω scans PANalytical X'Pert Pro MRD using Cu Kα radiation) as explained in detail in Ref.[4]. The In content from SIMS was obtained using a concentration calibration.[4] The sample surface was investigated by atomic force microscopy (AFM Bruker Dimension Edge) in PeakForce tapping mode.

| $T_g$ [°C] | O-flux [sccm] | Offcut [°] | Growth rate [nm/min] | | In content [cm$^{-3}$] |
|---|---|---|---|---|---|
| 800 | 0.75 | 0 on ($\bar{2}$01) (Ref.[4]) | 2.2 | (SIMS) | $1 \times 10^{19}$ |
| | | 0 | 0.1 | (SIMS) | |
| | | 2 | 0.17 | (SIMS) | |
| | | 6 (from Ref.[4]) | 0.33 | (SIMS) | $2 \times 10^{18}$ |
| 800 | 1 | 2 | 0.17 | (SIMS) | |
| | | 4 | 0.33 | (XRD) | |
| | | 6 | 0.47 | (XRD + SIMS) | $2 \times 10^{18}$ |
| 740 | 0.75 | 0 | 0.17 | (SIMS) | |
| | | 2 | 0.2 | (SIMS) | |
| | | 4 | 0.37 | (XRD) | |
| | | 6 | 0.8 | (XRD + SIMS) | $3\text{-}7 \times 10^{18}$ |
| 700 | 0.75 | 0 | 0.33 | (SIMS) | $2 \times 10^{19}$ |
| | | 2 | 0.33 | (XRD + SIMS) | $1.1 \times 10^{19}$ |
| | | 4 | 0.67 | (XRD) | |
| | | 6 | 1.5 | (XRD + SIMS) | $1.6\text{-}1.9 \times 10^{19}$ |

**Table 1** Growth rates of (100)-oriented β-Ga$_2$O$_3$ layers deposited in four different growth runs via MEXCAT-MBE (different offcuts co-loaded). A growth rate for ($\bar{2}$01) homoepitaxy is also reported for comparison (data taken from Ref.[4]). In the growth rate column we report in parenthesis the experimental methods used to determine the thickness. The In-content obtained from SIMS is reported for samples ≥ 10 nm.

In Table 1 we report a summary of the collected results. In agreement with our previous investigation of the Ga$_2$O$_3$ growth kinetics and thermodynamics by MBE and MEXCAT-MBE a lower $T_g$ or higher O-flux increase GR,[4,13–15,20,22] but also increase the tendency of indium-incorporation during MEXCAT-MBE.[4] More importantly, for all the tested growth conditions the GR is generally higher for larger offcuts, as also shown in Figure 1(b). Moreover, comparing the lowest tested $T_g$ runs (*i.e.*, 740°C and 700°C) it is interesting to highlight that both the depositions on 0°- and 2°-off substrates show similar (740°C) or equal (700°C) GRs, while higher offcuts in both cases result in a clear GR increment [Table 1 and Figure 1(b)]. In the following we try to give a physical explanation for this result: as schematically illustrated in Figure 1 (a), an offcut along the $[00\bar{1}]$ direction in (100) β-$Ga_2O_3$ substrates results in the formation of (100) terraces with different widths – *i.e.*, 16.9 nm, 8.5 nm, and 5.6 nm for 2°-, 4°-, and 6°-off, respectively; 0.59 nm steps are formed among the terraces, leaving exposed $(\bar{2}01)$ oriented surfaces at the step edges[7] [Figure 1(a)].

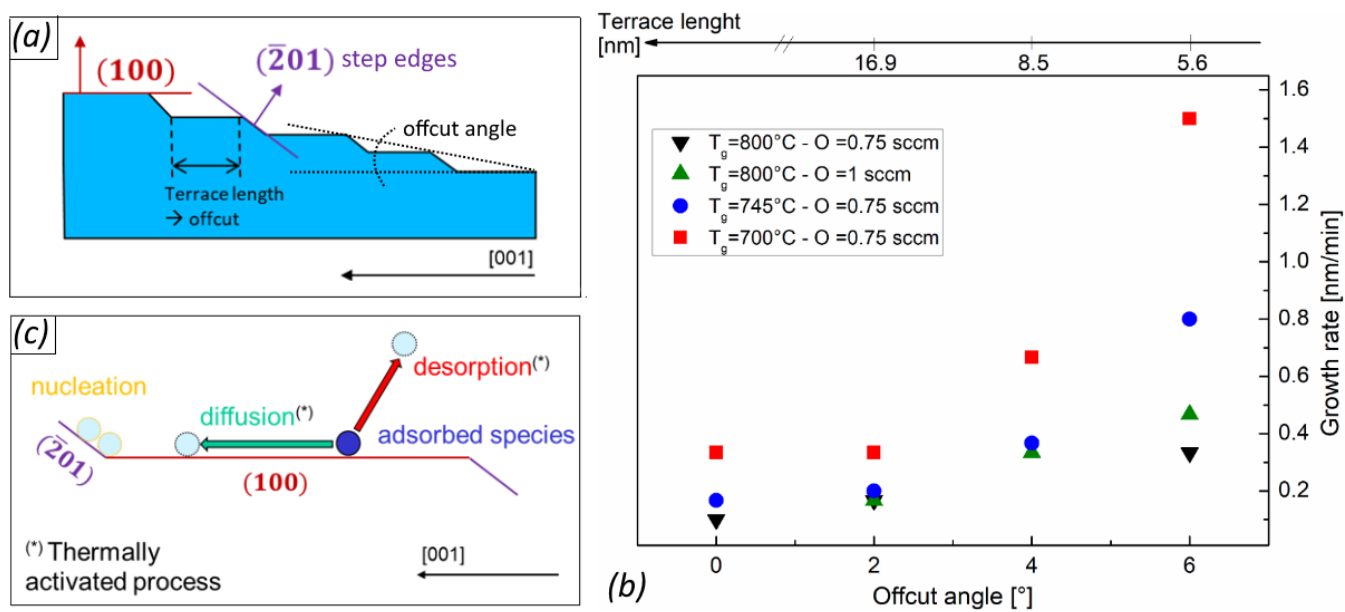

**Figure 1** (a) Sketch representing a cross-sectional view of (100) offcut substrates; (b) offcut and associated terrace length as a function of the obtained growth rate for different synthesis conditions (see Table 1); (c) schematic representation of the growth process on (100) offcut substrates via MBE.

The MBE growth on the (100) surface is complicated because of its high stability (*i.e.*, low surface free energy, leading to increased desorption during growth),[7,16] resulting in limited GRs;[3,18] even if mitigated, this is also true in the case of MEXCAT-MBE as proved by the data reported in Table 1 for 0°-off substrates. Furthermore, we have recently demonstrated[4] that the homoepitaxial growth on $(\bar{2}01)$-oriented substrates via MEXCAT-MBE allows for layer depositions at higher $T_g$ (and/or lower O-fluxes) with respect to the (100) orientation – *i.e.*, because of the higher surface energy of the $(\bar{2}01)$ with respect to the (100) one. As an example in Table 1 we give the reference growth rate measured on $(\bar{2}01)$-oriented β-$Ga_2O_3$ substrates (nominally 0°-offcut)[4] for the highest $T_g$/lowest O-flux investigated in the present study, which is significantly higher (2.2 nm/min) than that on 0°-offcut (100) substrate (0.1 nm/min).

Generally, growth takes place by a sequence of adsorption, diffusion, and nucleation competing with the desorption of the surface diffusing species.[26] In the particular case of oxides like $Ga_2O_3$ and $In_2O_3$[14] (both potentially involved at different stages of the MEXCAT-MBE deposition process[15,20]) their peculiar growth kinetics involves the intermediate formation of their respective volatile suboxide molecules ($Ga_2O$ and $In_2O$), which we consider the relevant, diffusing and desorbing species. We propose that a possible explanation of our experimental findings [Table 1, Figure 1(b)] can be found in *(i)* the ($\bar{2}01$) oriented step edges acting as nucleation sites for the surface diffusing species during growth [see Figure 1(c)] and *(ii)* the influence of growth condition on the diffusion length and desorption of the adsorbed species on the (100) surface. In order to reach the ($\bar{2}01$) edges, the adsorbed species must diffuse on the (100) terraces along the [001] direction and thus only species adsorbed within a diffusion length from the step edge can contribute to growth. Consequently, for the tested growth conditions ($T_g$ and O-fluxes) of this work, the diffusion length on the (100) surface along the [001] direction should be limited to less than the associated terrace length of the 4° offcut substrate, *i.e.*, ≈ 8.5 nm, to explain the growth-rate increase upon increasing the offcut to 6°, for example. Larger offcuts result in shorter terrace lengths and thus species adsorbed on a larger area fraction of the terraces are within a diffusion length from the step edges and can contribute to growth. Consequently, the GR increases with increasing offcut [Figure 1(b)]. The surface diffusion length $\lambda_S$ can be expressed as a function of the diffusion coefficient on the surface $D_S$ and the mean diffusion time $\tau_S$ as $\lambda_S = \sqrt{D_S \tau_S}$.[27] Importantly, $\tau_S$ is the time between adsorption and desorption of the diffusing species from the (100) terrace, *i.e.*, the inverse of the desorption rate. Since diffusion and desorption are thermally activated processes, $\lambda_S$ can be also expressed as a function of the barrier energy for the desorption rate and the surface diffusion coefficient $E_{des}$ and $E_{diff}$, respectively: $\lambda_S \propto e^{\frac{E_{des} - E_{diff}}{2kT_g}}$, where *k* is the Boltzmann constant. The decreasing GR with increasing $T_g$ [Figure 1(b)] at all chosen offcut angles indicates a decreasing $\lambda_S$, suggesting that $E_{des} > E_{diff}$ in our case, *i.e.*, a stronger increase of the desorption rate than increase of the surface diffusion coefficient.

A recent report on high temperature low pressure chemical vapour (HT-LPCVD) deposition of ($\bar{2}01$) β-$Ga_2O_3$ layers on offcut c-plane sapphire substrates also found an increasing GR with increasing offcut angle,[28] whereas MOVPE (100) homoepitaxy did not highlight any GR dependence for the same offcuts investigated in the present work.[8] These HT-LPCVD and MOVPE results are difficult to compare among each other due to different synthesis techniques, deposition conditions, and layer orientations; they however seem to indicate non-negligible and negligible desorption from the terraces, respectively.

The explanation presented here is also supported by the collected AFM micrographs obtained for the samples deposited at the lowest $T_g$ (700 °C, Figure 2). In such deposition conditions we report the growth of a ≈ 10 nm thick film in both 0°- and 2°-off substrates [XRD trace in Figure 2(b) (green) is in agreement with results obtained from SIMS (not shown)]. Nonetheless, in both samples it is possible to identify the presence of islands elongated along the [010] direction [Figure 2 (a), 0° and 2°]. Such islands have already been highlighted in homoepitaxial layers grown by MOVPE[29] and MBE[10] on exactly oriented (100) substrates and their elongation could be a sign of a lower diffusion length of the adsorbed species on the (100) surface along the [001] direction than in the [010] one. The coalescence of these islands usually results in the formation of twins.[9] The same could potentially happen also for the layers deposited on top of the 2°-offcut substrate [Figure 2(a)] in our MEXCAT-MBE growth due to islands nucleation on the terraces of the 2°-off substrate [visible as the interruption of the regular step-terrace lines along the [010] direction in Figure 2(a)] instead of nucleation at the ($\bar{2}$01) step edges. Differently, the surface morphology of the 4° and 6°-off samples [Figure 2(a)], despite larger film thicknesses with respect to the 0° and 2°-off ones [Figure 2(b)], suggests a step-flow or (partially) a step-bunched growth resulting in a more regular and smooth layer surface.

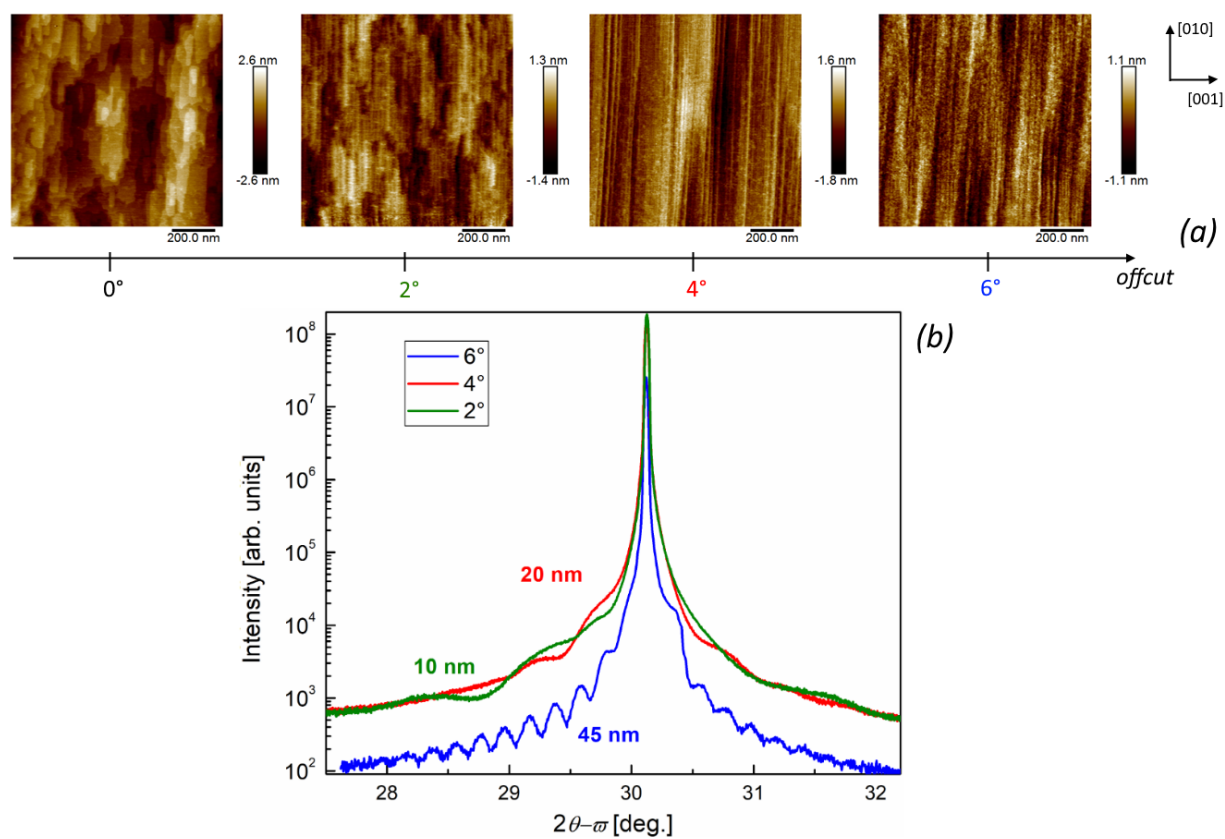

**Figure 2** (a) AFM micrographs of (100) β-Ga$_2$O$_3$ homoepitaxial layers deposited via MEXCAT-MBE at T$_g$ = 700°C with O-flux = 0.75 sccm on different offcut substrates [red squares in Figure 1(b)]. In (b) the corresponding XRD scans of the (400) reflection for the 2°, 4°, and 6°-off samples are reported with relative thicknesses from respective thickness fringes.

We now focus on the 6°-off samples deposited under the 4 different tested growth conditions (Table 1). Their AFM images [Figure 3 (a)] show the effect of the substrate temperature and (just for one sample deposited at 800 °C) of the O-flux on the surface morphology. Under all the deposition conditions for the 6°-off layers no structural defects are expected, as already evidenced by the TEM investigation of the sample deposited at 800 °C with O-flux = 0.75 sccm presented in Ref.[4]. Nonetheless, despite the limited root mean square roughness (rms = 0.3 nm) of this layer [Figure 3 (a)], step bunching can be highlighted (see TEM of this layer in Figure 5 of reference[4]). An increase in the O-flux while maintaining the very same $T_g$ resulted in an increased step bunching mechanism (rms = 1.3 nm) probably related to an increased diffusion length by suppressed desorption[14] through the increased oxygen species on the surface. On the other hand, decreasing the $T_g$ down to 740 °C while maintaining the O-flux at 0.75 sccm drastically reduced the step bunching, going towards a step-flow growth mechanism (rms = 0.19 nm). Nonetheless, the thickest layer deposited at the lowest $T_g$ (700 °C, O-flux = 0.75 sccm) shows an intermediate roughness (rms = 0.32 nm); the reasons are still not fully understood.

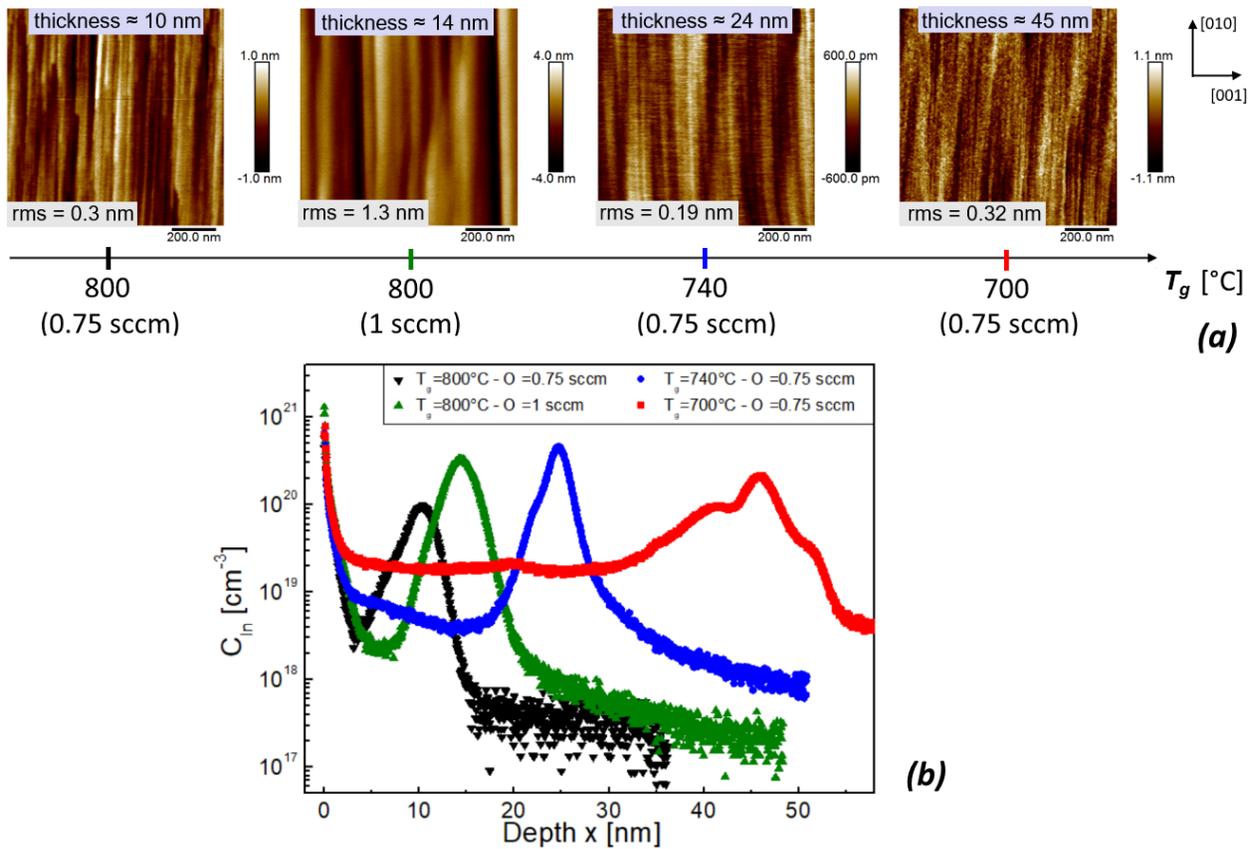

**Figure 3** (a) AFM micrographs of (100) β-$Ga_2O_3$ homoepitaxial layers deposited via MEXCAT-MBE at different $T_g$ or O-fluxes for the same 6° substrate offcut. (b) Indium concentration $C_{In}$ depth profiles for the same samples extracted by ToF-SIMS. The film thickness is equated with the depth of the In-accumulation peak at the layer-substrate interface. The AFM as well as the SIMS profile of the sample deposited at 800 °C with 0.75 sccm flux have already been presented in Ref.[4].

Being these samples deposited with In-mediated MEXCAT-MBE, both $T_g$ and O-flux employed decisively affect the concentration of incorporated In inside the deposited layer: a lower $T_g$ and/or a higher O-flux during the deposition process was shown to be connected to a larger amount of In inside the sample.[4] As visible from Figure 3 (b) (values also reported in Table 1) this trend is also confirmed for the (100) samples investigated in this work. Coherently, the deposition conditions that resulted in the highest growth rate of 1.5 nm/min – *i.e.*, $T_g$ = 700 °C, O-flux = 0.75 sccm – also result in the highest In-incorporation ($\approx$ 1.6-1.9 × 10$^{19}$ cm$^{-3}$). Nevertheless, as already discussed in a previous work[4] such concentrations should not affect the electrical properties of the layer as In is isovalent with Ga and such low concentrations would only lead to a negligible bandgap decrease. Moreover, the data reported in Table 1 on the series of samples deposited on the different offcut substrates under the same synthesis conditions (*i.e.*, $T_g$ = 700 °C, O-flux = 0.75 sccm – 0°, 2°, and 6°) consistently show that the amount of incorporated In is not affected by the offcut itself, but just by the deposition conditions.

In conclusion, we have shown how In-mediated MEXCAT MBE can be considered as a viable deposition technique for the deposition of high quality (100) β-Ga$_2$O$_3$ homoepitaxial layers. In particular, we have demonstrated step-flow growth on substrates offcut towards the [00$\bar{1}$] direction at growth rates comparable to the ones obtained by MOVPE technique on this substrate orientation. Different from MOVPE results,[8] during MEXCAT-MBE an increasing growth rate with increasing offcut angle was found and related to almost exclusive layer nucleation on the ($\bar{2}$01)-oriented step edges due to their higher surface free energy (and thus lower propensity for desorption) than that of the (100)-oriented terraces. Regarding the obtained absolute growth rate value of 1.5 nm/min the authors remind that this could be scaled by scaling up the Ga and O-fluxes as already demonstrated for the (010) orientation.[5] We believe that this result, together with the deep understanding of the underlying physical processes could represent an important step further for the realization of β-Ga$_2$O$_3$-based power electronic heterostructured and/or homostructured devices by MBE on (100)-oriented substrates.

**Acknowledgments:**

We would like to thank Jens Herfort for critically reading the manuscript, as well as Hans-Peter Schönherr, Carsten Stemmler, and Katrin Morengroth for technical MBE support. This work was performed in the framework of GraFOx, a Leibniz-Science Campus partially funded by the Leibniz Association.

**References:**

[1] S.J. Pearton, J. Yang, P.H. Cary, F. Ren, J. Kim, M.J. Tadjer, and M.A. Mastro, Appl. Phys. Rev. **5**, 011301 (2018).

[2] Z. Galazka, Semicond. Sci. Technol. **33**, 113001 (2018).

[3] K. Sasaki, A. Kuramata, T. Masui, E.G. Víllora, K. Shimamura, and S. Yamakoshi, Appl. Phys. Express **5**, 035502 (2012).

[4] P. Mazzolini, A. Falkenstein, C. Wouters, R. Schewski, T. Markurt, Z. Galazka, M. Martin, M. Albrecht, and O. Bierwagen, APL Mater. **8**, 011107 (2020).

[5] A. Mauze, Y. Zhang, T. Itoh, F. Wu, and J.S. Speck, APL Mater. **8**, 021104 (2020).

[6] Y. Zhang, F. Alema, A. Mauze, O.S. Koksaldi, R. Miller, A. Osinsky, and J.S. Speck, APL Mater. **7**, 022506 (2018).

[7] R. Schewski, K. Lion, A. Fiedler, C. Wouters, A. Popp, S.V. Levchenko, T. Schulz, M. Schmidbauer, S. Bin Anooz, R. Grüneberg, Z. Galazka, G. Wagner, K. Irmscher, M. Scheffler, C. Draxl, and M. Albrecht, APL Mater. **7**, 022515 (2018).

[8] S. Bin Anooz, R. Grüneberg, C. Wouters, R. Schewski, M. Albrecht, A. Fiedler, K. Irmscher, Z. Galazka, W. Miller, G. Wagner, J. Schwarzkopf, and A. Popp, Appl. Phys. Lett. **116**, 182106 (2020).

[9] G. Wagner, M. Baldini, D. Gogova, M. Schmidbauer, R. Schewski, M. Albrecht, Z. Galazka, D. Klimm, and R. Fornari, Phys. Status Solidi A **211**, 27 (2014).

[10] Z. Cheng, M. Hanke, Z. Galazka, and A. Trampert, Nanotechnology **29**, 395705 (2018).

[11] A. Fiedler, R. Schewski, M. Baldini, Z. Galazka, G. Wagner, M. Albrecht, and K. Irmscher, J. Appl. Phys. **122**, 165701 (2017).

[12] Y. Oshima, E. Ahmadi, S. Kaun, F. Wu, and J.S. Speck, Semicond. Sci. Technol. **33**, 015013 (2018).

[13] P. Vogt and O. Bierwagen, Appl. Phys. Lett. **108**, 072101 (2016).

[14] P. Vogt and O. Bierwagen, Phys. Rev. Mater. **2**, (2018).

[15] O. Bierwagen, P. Vogt, and P. Mazzolini, in *Gallium Oxide Mater. Prop. Cryst. Growth Devices*, edited by M. Higashiwaki and S. Fujita (Springer International Publishing, Cham, 2020), pp. 95–121.

[16] V.M. Bermudez, Chem. Phys. **323**, 193 (2006).

[17] E. Ahmadi, O.S. Koksaldi, S.W. Kaun, Y. Oshima, D.B. Short, U.K. Mishra, and J.S. Speck, Appl. Phys. Express **10**, 041102 (2017).

[18] M.-Y. Tsai, O. Bierwagen, M.E. White, and J.S. Speck, J. Vac. Sci. Technol. A **28**, 354 (2010).

[19] P. Mazzolini and O. Bierwagen, J. Phys. Appl. Phys. **53**, 354003 (2020).

[20] P. Vogt, O. Brandt, H. Riechert, J. Lähnemann, and O. Bierwagen, Phys. Rev. Lett. **119**, 196001 (2017).

[21] M. Kracht, A. Karg, J. Schörmann, M. Weinhold, D. Zink, F. Michel, M. Rohnke, M. Schowalter, B. Gerken, A. Rosenauer, P.J. Klar, J. Janek, and M. Eickhoff, Phys Rev Appl. **8**, 054002 (2017).

[22] P. Mazzolini, P. Vogt, R. Schewski, C. Wouters, M. Albrecht, and O. Bierwagen, APL Mater. **7**, 022511 (2018).

[23] P. Vogt, A. Mauze, F. Wu, B. Bonef, and J.S. Speck, Appl. Phys. Express **11**, 115503 (2018).

[24] Z. Galazka, R. Uecker, D. Klimm, K. Irmscher, M. Naumann, M. Pietsch, A. Kwasniewski, R. Bertram, S. Ganschow, and M. Bickermann, ECS J. Solid State Sci. Technol. **6**, Q3007 (2017).

[25] Z. Galazka, K. Irmscher, R. Uecker, R. Bertram, M. Pietsch, A. Kwasniewski, M. Naumann, T. Schulz, R. Schewski, D. Klimm, and M. Bickermann, J. Cryst. Growth **404**, 184 (2014).

[26] W. Miller, D. Meiling, R. Schewski, A. Popp, S.B. Anooz, and M. Albrecht, Phys. Rev. Res. **2**, 033170 (2020).

[27] I. Bryan, Z. Bryan, S. Mita, A. Rice, J. Tweedie, R. Collazo, and Z. Sitar, J. Cryst. Growth **438**, 81 (2016).

[28] Y. Zhang, Z. Feng, M.R. Karim, and H. Zhao, J. Vac. Sci. Technol. A **38**, 050806 (2020).

[29] R. Schewski, M. Baldini, K. Irmscher, A. Fiedler, T. Markurt, B. Neuschulz, T. Remmele, T. Schulz, G. Wagner, Z. Galazka, and M. Albrecht, J. Appl. Phys. **120**, 225308 (2016).